\documentclass{caosp305}

\usepackage{graphicx}
\usepackage{natbib}
\articleNo{123}
\pubyear{2018}
\volume{35}
\volnumber{3}
\firstpage{1}
\received{November 10, 2017}
\accepted{November 10, 2017}

\def\BibTeX{{\rm B\kern-.05em{\sc i\kern-.025em b}\kern-.08em
             T\kern-.1667em\lower.7ex\hbox{E}\kern-.125emX}}

\begin{document}

\htitle{Stars with a Stable Magnetic Field}
\hauthor{Lilia Ferrario}

\title{Stars with a Stable Magnetic Field}

\author{
        Lilia Ferrario
       }

\institute{
           Mathematical Sciences Institute\\
           The Australian National University\\
           ACT 2601, Australia, \email{Lilia.Ferrario@anu.edu.au}\\
          }

\date{November 10, 2017}

\maketitle

\begin{abstract}
  In this review I will summarise what we know about magnetic fields
  in stars and what the origin of these magnetic fields may
  be. I will address the issue of whether the magnetic flux is
  conserved from pre-main sequence to the compact star phase (fossil
  field origin) or whether fields may be dynamo generated during some
  stages of stellar evolution or perhaps during stellar merging
  events.

  \keywords{stars: magnetic field -- stars: protostars -- stars:
    chemically peculiar -- white dwarfs -- pulsars -- stars:
    magnetars}

\end{abstract}

\section{Introduction} \label{intr}

\begin{figure}
\centerline{\includegraphics[width=0.75\textwidth,clip=]{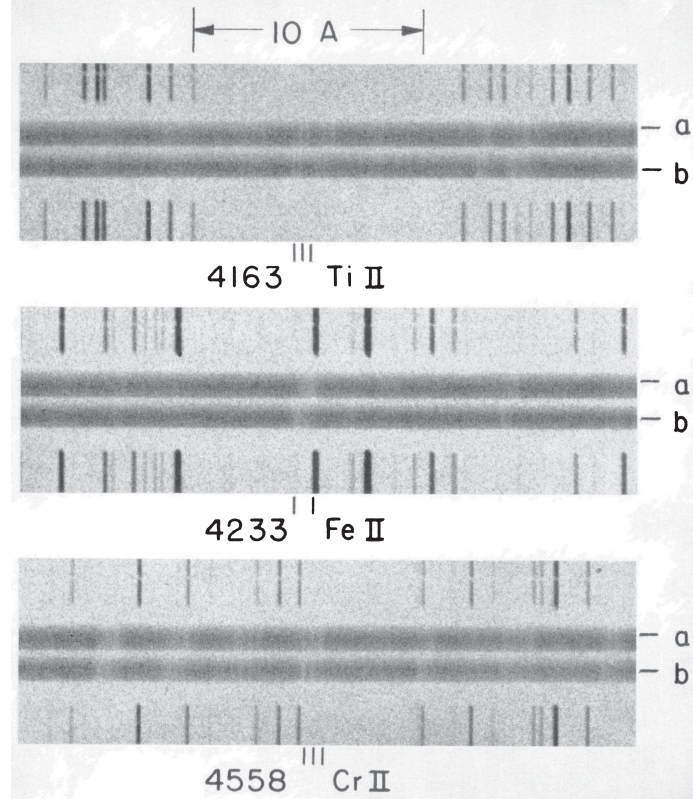}}
\caption{Spectrum of HD\,215441 \citep{Babcock1947} showing Zeeman
  split lines in a magnetic field of about 34\,000\,G. The spectra marked by
  $a$ and $b$ were photographed simultaneously through the left-hand
  and right-hand sections of a double circular analyzer. The original
  dispersion 4.5\,\AA\,mm$^{-1}$; slit width corresponds to
  0.14\,\AA.}
\label{BabcockStar}
\end{figure}

\citet{Babcock1947} was the first astronomer to detect a magnetic
field in a star (78 Vir). He also discovered what is still now the
most highly magnetic main sequence star, HD\,215441 whose field
strength is about $3.4\times 10^4$\,G \citep{Babcock1960}. This became
known as ``Babcock's star'' (see Fig.\,\ref{BabcockStar}).

The existence of strong magnetic fields in white dwarfs was revealed
much later when \citet{Kempetal1970} detected strong circular
polarisation in the continuum of Grw$+70^{\circ}8247$ while in neutron
stars came with the discovery of the first pulsar, a fast spinning
neutron star \citep{Pacini1968}, by Jocelyn Bell in 1967
\citep{Hewish1968}.

Since those early years, major progress has been made on stellar
magnetism thanks to high quality data coming from surveys covering the
full range of stellar masses and spanning all evolutionary phases,
including pre-main sequence
\citep[e.g.][]{Alecian2013a,Alecian2013b,HubrigIlyin2013}, main and
post main-sequence \citep[e.g.][]{Hubrig2011, Auriere2015, Wade2016,
  Mathys2017}, white dwarfs \citep[e.g.,][]{Schmidt2001,
  Landstreet2012, Kepler2013} and neutron stars \citep[the Parkes
multibeam pulsar survey;][]{Manchester2001}. The data secured by these
surveys have allowed researchers to explore the incidence of magnetism
among stars, probe their magnetic field strength and structure, and
study their field evolution and origin.

The origin of large scale and stable magnetic fields in stars remains
an open question in astrophysics.  In this paper I will review what is
currently known about stellar magnetism and the hypotheses that have
been advanced to explain the origin of magnetic fields in stars. A
comprehensive paper on stable magnetic equilibria and their evolution
in main-sequence and compact stars can be found in
\citet{Reisenegger2009} and an extensive review on the origin of
stellar magnetic fields in \citet{Ferrario2015Origin}.

\section{The Dichotomies of Magnetic Non-Degenerate Stars}

Herbig Ae/Be (HAeBe) objects are pre-main sequence stars of
$2-15$\,M$_\odot$ at the later stages of their formation.  These
objects are still immersed in their proto-stellar gas-dust envelope
and exhibit emission lines of stellar type A/B.
\citet{Alecian2013a,Alecian2013b} and \citet{HubrigIlyin2013}
conducted large spectro-polarimetric surveys of HAeBe stars to
investigate their magnetic and rotational properties and explore
possible evolutionary links between their characteristics to those of
the magnetic main sequence Ap/Bp stars. These observations have indeed
shown that around 7\% of HAeBe objects display large-scale, mainly
dipolar magnetic fields with strengths in the range $300-2\,100$\,G
which are similar to those of magnetic main sequence Ap and Bp stars
if conservation of magnetic flux is assumed. The incidence of
magnetism among HAeBe is also consistent with that observed on the
main-sequence. The surveys have also revealed that while the
non-magnetic HAeBe objects exhibit rotations
$v\sin i=0-300$\,km\,s$^{-1}$, the magnetic ones are generally much
more slowly rotating with $v\sin i\le 100$\,km\,s$^{-1}$.  This
dichotomy is reminiscent of that observed in magnetic stars on the
main sequence \citep[as first reported by][]{Wolff1975} that can have
spin periods of up to 1\,000 years \citep[see][]{Mathys2015}. This
suggests that those physical processes that are responsible for
rotational braking in magnetic main-sequence stars are already at play
in the late stages of stellar formation. In this context,
\citet{Netopil2017} and Netopil (2018, these proceedings) have studied
the rotational characteristics of a sample of more than 500 magnetic
main-sequence stars. Their results have confirmed the previous results
of \citet{North1998} and \citet{Stepien1998} that the angular momentum
is conserved during the evolution on the main-sequence and that no
further magnetic braking is observed. Furthermore, they showed that
while stars with the highest fields tend to be the slowest rotators,
the strongest fields are found only in stars with spin periods shorter
than about 150\,d, thus confirming the earlier findings of
\citet{Mathys1997} and more recently of \citet{Mathys2017}.

\begin{figure}
\centerline{\includegraphics[width=0.75\textwidth,clip=]{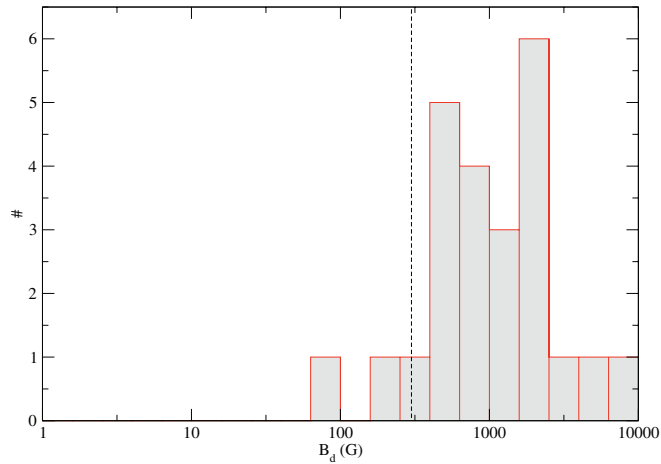}}
\caption{Histogram of the derived dipole strengths $B_d$ of a sample
  of 28 magnetic Ap/Bp stars from \citet{Auriere2007}. Note the dearth
  of Ap stars with $B_d<300$\,G.}
\label{Auriere_dichotomy}
\end{figure}

A second dichotomy exists among A and B type stars with masses of
$1.5-6$\,M$_\odot$, that is, they either exhibit large-scale fields of
$300-34\,000$\,G or they are not magnetic down to the current
detection limit of a few Gauss (see Fig\,\ref{Auriere_dichotomy}). The
lack of stars in the 1-300\,G field regime has been called the ``Ap/Bp
magnetic desert'' \citep{Auriere2007}. Below the magnetic desert lies
another type of magnetic stars, typified by Vega, whose longitudinal
field is at the sub-Gauss level \citep{Lignieres2009}. Thus this
magnetic desert separates intermediate-mass stars with large-scale
stable fields from those with unstable fields suggesting that two
different mechanisms may be responsible for the generation of their
fields. However, \citet{Fossati2015} have reported that massive stars
may not have a magnetic desert, although its absence may be linked to
mass-dependent field decay \citep[see below][]{Fossati2016}.

A third dichotomy concerns the dearth of close binaries among
main-sequence magnetic stars.  The aim of the BinaMIcS (Binarity and
Magnetic Interactions in various classes of Stars) programme
\citep{Alecian2015} was initiated to explore the incidence of
magnetism in binaries with periods shorter than 20\,d. Their studies
have shown an incidence of magnetism that is about 3 to 5 times lower
than in non-magnetic stars, thus confirming the results of
\citet{Carrier2002} who first reported that binary systems hosting an
Ap star tend to have periods $\ge 3$\,d.

The above three dichotomies tell us the following. It appears that
large-scale magnetic fields in proto-stars and on the main-sequence
are present only in a small fraction ($\sim 7$\%) of stars with
radiative envelopes and that these magnetic stars are very rarely
found in close binaries. The magnetic field strength dichotomy tells
us that the origin of stellar magnetism cannot be attributed to some
dynamo action taking place in some evolutionary phases, because if
this were the case all stars would be magnetic at some level and there
would be no Ap/Bp magnetic desert.

Observations indicate that the magnetic flux is conserved during the
evolution from the pre-main sequence to the main sequence. However, if
magnetism is a relic of the interstellar field from which the stars
formed \citep[see, e.g.,][]{Mestel1966} then we would expect that the
incidence of magnetism should vary across diverse stellar
populations. Thus the incidence in the Galactic field should differ
from that in clusters and also from cluster to cluster. However, this
does not seem to be the case \citep[e.g.][]{Paunzen2005, Paunzen2006}.
It is also curious that all magnetic stars in binaries \citep[with the
exception of $\epsilon$\,Lupi;][]{Shultz2015} have non-magnetic
companions.
\begin{figure}
\centerline{\includegraphics[width=0.75\textwidth,clip=]{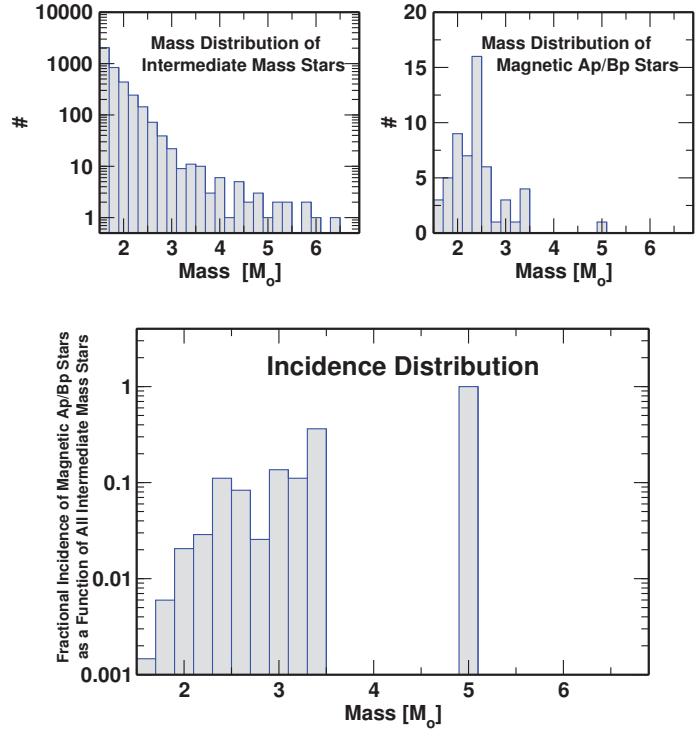}}
\caption{Mass distribution of A/B stars in the solar
  neighbourhood. Top left: mass distribution of all intermediate mass
  A/B stars. Top right: mass distribution of magnetic Ap/Bp
  stars. Bottom: incidence of Ap/Bp stars as a function of mass
  \citep{Power2007}.}
\label{PowerJenniferVolumeLimited}
\end{figure}

In order to explain the above dichotomies, \citet{Ferrario2009}
advanced the hypothesis that magnetic fields could form when two
proto-stars merge as they approach the main-sequence and when at least
one of them has already developed a radiative envelope.  These late
mergers would produce a brief period of strong differential rotation
and give rise to the large-scale fields observed in the radiative
envelopes of Ap, Bp, and Of?p stars. They would also explain the
scarcity of close binaries among intermediate-mass main-sequence
magnetic stars. One of the merging proto-star predictions is that the
incidence of magnetism should grow with stellar mass, which appears to be
validated by the studies of \citet{Power2007} (see
Fig.\,\ref{PowerJenniferVolumeLimited}) and Sikora et al. (2018, these
proceedings) who conducted a volume-limited sample of A/B stars with
$M\le 4$\,M$_\odot$.

The question of how fields evolve with stellar age was addressed by
\citet{Landstreet2008} who performed surveys of magnetism in A and B
type stars in clusters.  These observations allowed them to link field
strength and structure to stellar mass, fractional ages, and
metallicity. Their work revealed that fluxes clearly decrease with age
indicating field decay. They also reported that such a decline is
faster in stars with a mass larger than about
3\,M$_\odot$. \citet{Fossati2016} conducted a similar study on the
field evolution in massive (5-100\,M$_\odot$) main-sequence stars and
found an obvious deficiency of evolved magnetic stars which is more
prominent at higher masses. So they propose that the absence of a
magnetic desert in massive magnetic stars may be caused by the
mass-dependent timescales over which field decay occurs. That is, if
all magnetic stars are born with a field above a certain cutoff, the
lowest mass stars would retain it as they evolve on the
main-sequence. This is because the field decay times in these stars is
comparable to or longer than their main-sequence lifetime. However, if
field decay acts faster as mass increases (and thus on timescales that
are progressively shorter than their main-sequence lifetime), then
massive magnetic stars will display, as a class, a smooth magnetic
field distribution that extends to very low fields.

\citet{Fossati2016} also noted that field decay may explain the
intriguing mismatch between the percentage of massive magnetic stars
and that of slowly rotating massive stars. That is, many massive stars
that used to be strongly magnetic (and therefore slowly rotating) are
no longer magnetic at any measurable level \citep[see
also][]{Fossati2015}.

Despite this decay, fields have been observed in some
post-main-sequence stars. The first field discovered in an evolved
star was that in EK\,Eridani \citep{Auriere2011}. This is a very
slowly rotating red giant ($P=308.9$\,d) with a convective envelope
and a large scale poloidal field of about 270\,G.  \citet{Neiner2017}
detected magnetic fields in another two hot evolved stars:
$\iota$\,Car (either on its first crossing of the HR diagram or on a
blue loop) and HR\,3890 (on its first and only crossing of the HR
diagram) and confirmed the field in $\epsilon$\,CMa (near the end of
the main-sequence) first reported by \citet{Fossati2015}. Their field
strengths are compatible with magnetic flux conservation during
stellar evolution indicating that at least in a fraction of stars
magnetic fields can persist when a star evolves off the main-sequence.

The magnetic Ap/Bp stars show periodic variabilities caused by the
non-uniform distribution of chemical elements on their surface. It is
still not clear what causes these inhomogeneities and the viability of
theoretical models is mostly restricted to observations of
Galactic objects. However, stellar magnetism data 
obtained in extragalactic systems with different environmental
properties would give us additional information that would allow us to
better constrain theoretical models. This is what motivated
\citet{Paunzen2013} to conduct photometric observations of a sample of
magnetic Ap/Bp candidates in the Large Magellanic Cloud (LMC). They
discovered that their LMC sample exhibits a low variability
amplitude and explained it as due to the absence of regions
of the stellar surface that have an overabundance of optically active
elements.  \citet{Paunzen2005,Paunzen2006} also found that while the
percentage of chemically peculiar stars in Galactic open clusters is
almost identical to that of the Galactic field, in the LMC the
incidence is only about 2.2\% -- less than half of that
estimated in the Galaxy. On the other hand, stellar masses and ages do
not seem to be dissimilar from Galactic objects. These authors note
that the LMC metallicity is about 0.5\,dex compared to the Sun thus
yielding important insights into the origin of chemical peculiarities
and on the origin of magnetic fields in stars.

There are currently no calculations that include fossil fields and
follow their evolution as the star ages and all the results
highlighted in this section provide us with valuable
constraints for the construction of evolutionary models.

A very comprehensive review of magnetic fields in non-degenerate stars is
given by \citet{DonatiLandstreet2009}.

\section{Magnetic White Dwarfs}

Magnetic fields of isolated magnetic white dwarfs lie in the range
$10^3-10^9$\,G. The upper limit cutoff near $10^9$\,G may be real but
the incidence of magnetism below a few $10^3$\,G still needs to be
established \citep[see][]{Landstreet2012, Landstreet2017b}.  Since the
discovery of the first magnetic white dwarf
\citep[Grw$+70^{\circ}8247$;][]{Kempetal1970} the number of objects
has been steadily increasing. We now have about 300 isolated and 170
magnetic white dwarfs in interacting binaries (the magnetic
cataclysmic variables, MCVs). Extensive reviews on magnetism in white
dwarfs can be found in \citet{Ferrario2015MWD}, \citet{WickFer2000}
and also Kawka (2018, these proceedings).

The origin of magnetic fields in white dwarfs has been vigorously
debated since their discovery. The proposal that the magnetic Ap/Bp
stars are the progenitors of the High Field Magnetic White Dwarfs
\citep[HFMWDs;][]{twf04,Wickramasinghe2005}, as first suggested by
\citet{Woltjer1960}, has recently been questioned \citep{Liebert2005,
  Liebert2015}. The point is that there should be the same fraction of
HFMWDs in binaries as in single stars.  The Sloan Digital Sky Survey
has identified thousands of detached white dwarf -- M dwarf
spectroscopic binaries \citep[e.g.,][]{Rebassa2013, Rebassa2016,
  Ferrario2012} but none of these has a field above a few $10^6$\,G
(the detectable limit in the SDSS spectra) even if there are hundreds
of HFMWDs known to have fields greater than this detection limit
\citep{Liebert2005, Liebert2015}.  The sample of white dwarfs within
20\,pc \citep{Holberg2008} has shown that $19.6\pm4.5$\% of white
dwarfs have main-sequence companions.  Thus, $14-24$\% of the
$\sim 300$ HFMWDs should also have such companions, but none has been
identified.  The magnitude-limited Palomar-Green survey has shown that
$23-29$\% of hot white dwarfs have cool companions.  Thus, we expect
$70−90$ of the $\sim 300$ HFMWDs to have a companion.  However, none
has been found.  The logical conclusion is that the origin of high
magnetic fields in white dwarfs is intimately related to their
binarity, as first proposed by \citet{Tout2008}.  We know that some
HFMWDs are the result of merging events \citep[EUVE\,0317-855 is
probably the best example,][]{Vennes2003}.  Thus, if magnetic fields
in white dwarfs arise as a result of either double degenerate or
common envelope mergers, then the complete absence of main sequence
(generally M-dwarf) companions to HFMWDs can easily be
explained. Despite the total absence of detached binaries composed by
a HFMWD and a non-degenerate companion, there are quite a few examples
of binaries comprising two white dwarfs, one of which is highly
magnetic \citep[see][and references therein]{Kawka2017}. Close
post-common envelope magnetic binaries could have developed their
fields during the common envelope phase \citep[e.g., the fast spinning
super-Chandrasekhar system NLTT\,12758]{Kawka2017}, while distant 
binaries could have been initially triple systems with two of the
three stars merging at some point (e.g., EUVE\,0317-855). Following
\citet{Tout2008}, the stellar merging hypothesis has been explored by
\citet{Bogomazov2009}, \citet{Nordhausetal2011},
\citet{Garciaberro2012} and \citet{WTF2014}.

One feature that characterises the HFMWDs is that their mean mass is
around $0.78$\,M$_\odot$, considerably larger than that of
non-magnetic white dwarfs
\citep[$0.66$\,M$_\odot$,][]{Tremblay2013}. The population synthesis
calculations of \citet{Briggs2015} used the incidence of magnetism
among white dwarfs and their mass distribution to constrain their
models computed for different values of the common envelope efficiency
parameter $\alpha$.  They found that the best agreement with
observations is obtained for $\alpha<0.3$ with the major contribution
coming from the merging of a late asymptotic giant branch star with a
main-sequence star.

A field distribution similar to that of the isolated HFMWDs is
observed in the MCVs. In these interacting systems the mass flowing
from the M-dwarf to the magnetic white dwarf is funnelled by the
strong fields \citep[a few $10^6-10^8$\,G, as revealed by photospheric
Zeeman lines and/or cyclotron emission features in their UV to IR
spectra, e.g.,][]{Ferrario1992, Ferrario1993a, Ferrario1996,
  Ferrario2003, Schwope2003, Hoard2004} to form accretion shocks near
the magnetic poles. Truncated accretion discs may or may not be
present depending on the accretion rate and field strength
\citep[e.g.,][]{Ferrario1993b, Ferrario1999}. The birth properties of
MCVs have been analysed in the context of the common envelope origin
of magnetic fields by Briggs et al. (2018, these
proceedings). According to this scenario, those systems that emerge
from common envelopes as close binaries and about to exchange mass will
evolve into MCVs. They found that the best agreement with observations
is obtained again for $\alpha<0.3$. The study by \citet{Zorotovic2010}
of the possible evolutionary histories of a sample of SDSS
post-common-envelope binaries is in good agreement with the Briggs et
al. (2018) results.

I note that there is an interesting parallel between the rarity of
close binaries among magnetic main-sequence stars and HFMWDs,
supporting a similar (merging) hypothesis for the origin of their
fields.

Magnetic white dwarfs are also becoming important tools to investigate the
formation and composition of exoplanets. For instance, the suggestion
that GD\,356 may have an Earth-type planetary companion \citep{Li1998}
is still a tantalising possibility to explain its unique emission line
spectrum \citep{Ferrario1997}. Furthermore
\citet{kawka2014} have shown that the incidence of magnetism among
cool and polluted white dwarfs is much greater than among their
non-magnetic counterparts, and proposed a link between crowded
planetary systems and the generation of magnetic fields in white
dwarfs (see also Kawka, 2018, these proceedings).

\section{Neutron Stars}

\begin{figure}
\begin{minipage}[c]{0.5\linewidth}
\includegraphics[width=\linewidth]{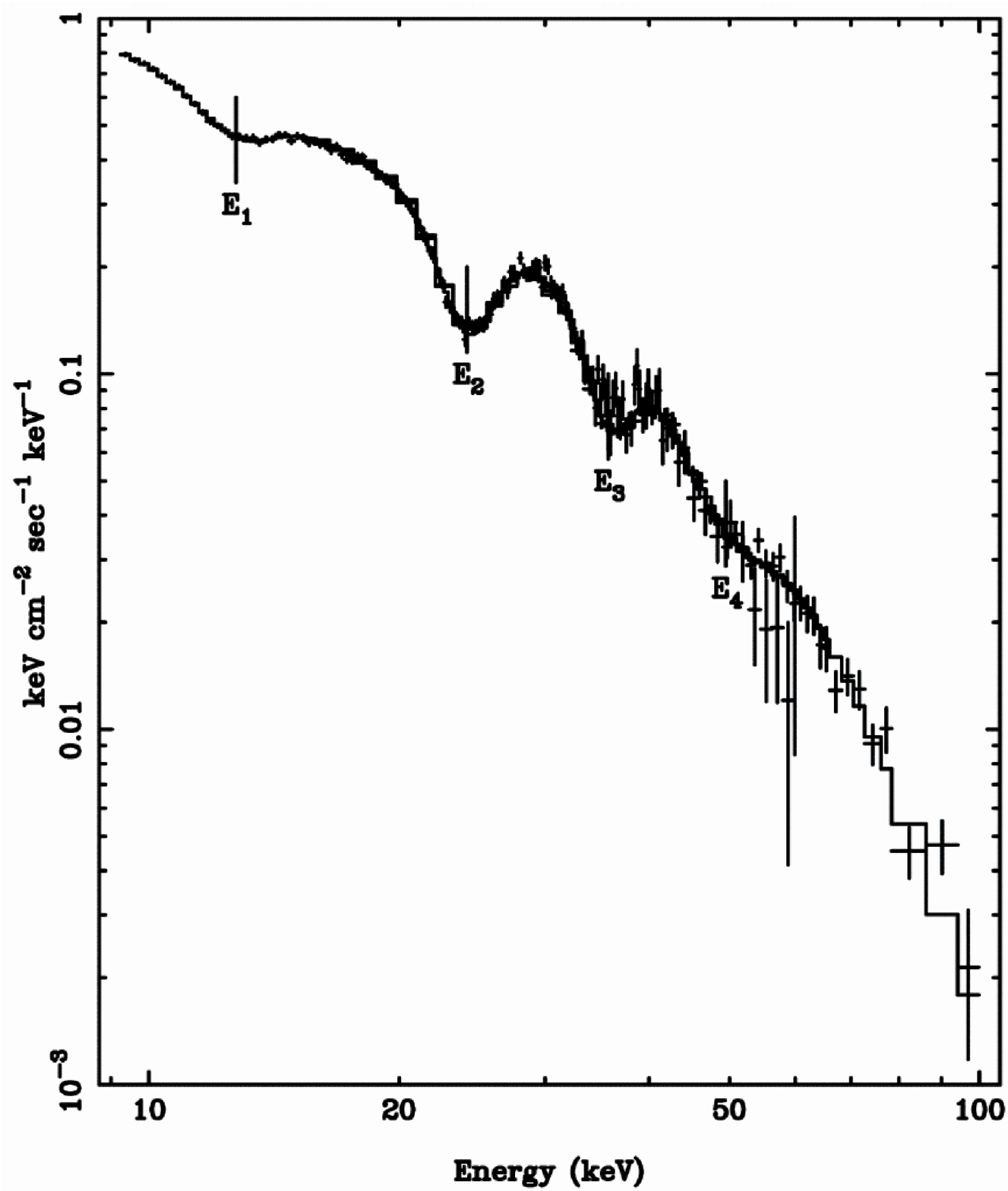}
\end{minipage}
\hfill
\begin{minipage}[c]{0.5\linewidth}
\includegraphics[width=\linewidth]{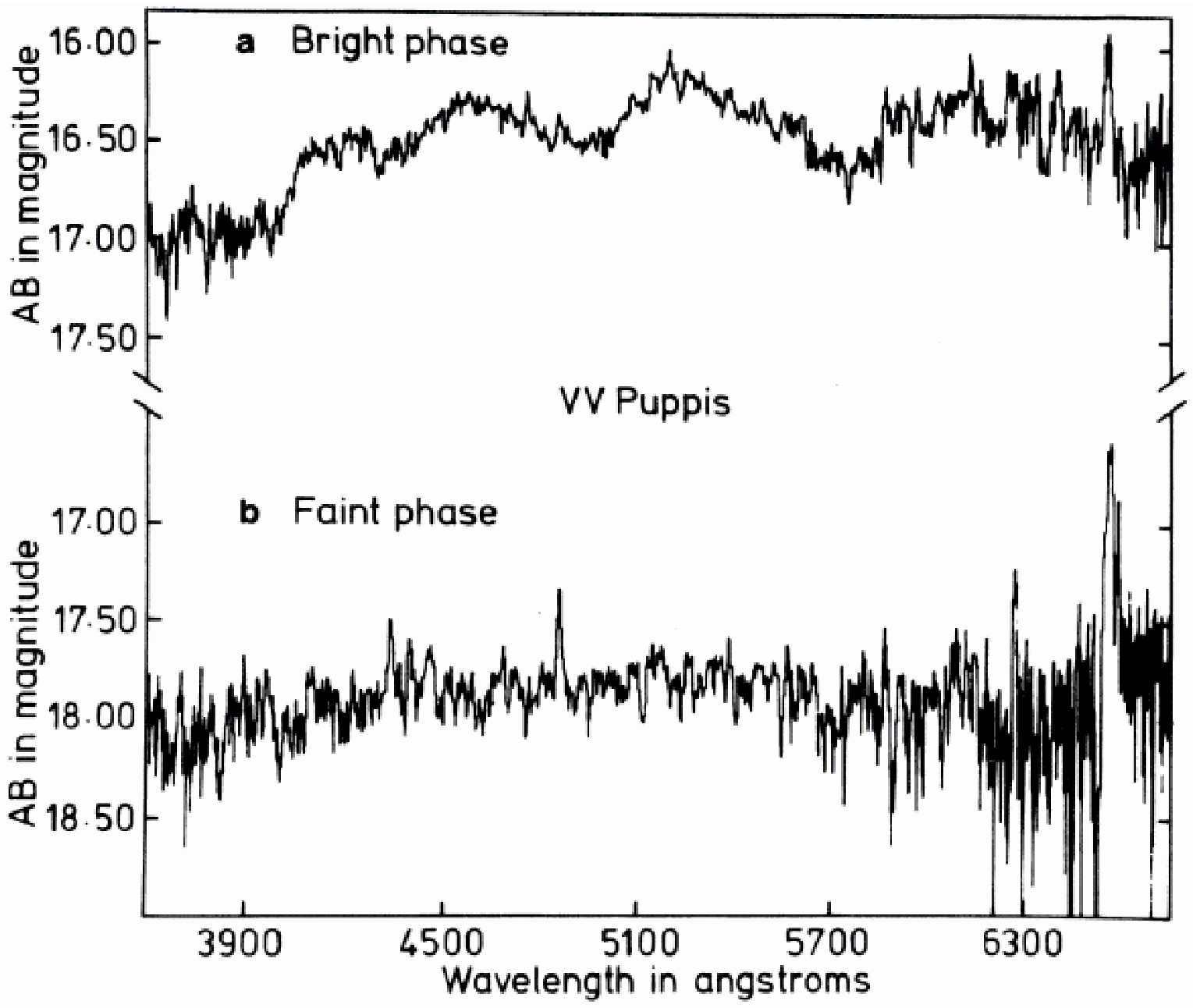}
\end{minipage}%
\caption{Left: Spectrum of X0115+63 showing cyclotron resonant
  scattering features suggestive that the accreting neutron star has a
  field of about $10^{12}\,$G \citep{Santangelo1999}. Right: The
  discovery spectrum of VV\,Puppis exhibiting cyclotron harmonic lines
  from the accretion shock on the surface of the magnetic white dwarf
  \citep{Vis1979} indicating a field strength of $3.2\times 10^7$\,G.}
\label{cyclotron_humps}
\end{figure}

The vast majority of neutron stars is made up by the classical radio
pulsars \citep[see the review by][]{Beskin2015Pulsars}. Pulsars are
powered by the loss of rotational energy caused by magnetic braking
\citep[remarkably,][suggested that a rotating pulsar might emit radio
waves just before their discovery]{Pacini1967}. If a pulsar has a spin
period $P$, typically $0.3-2$\,s, and a period derivative $\dot P$
(the rate at which the pulsar spins down), typically
$10^{-17}-10^{-13}$\,s\,s$^{-1}$, then the dipole radiation formula
$B=3.2\times 10^{19}\sqrt{P\dot P}$\,G \citep{Manchester1977} gives an
estimate of a few $10^{11}-10^{13}$\,G for the magnetic field
strength. Thus, in the vast majority of cases the magnetic field
strength of neutron stars can only be measured indirectly. The only
exception arises in those rare cases when the X-ray spectrum of
accreting neutron stars shows cyclotron harmonic features, such as
those observed in the accreting neutron stars in X0115+63
\citep{Santangelo1999}, Vela X-1 \citep{Kreykenbohm2002} and
1E1207.4-5209 \citep{Bignami2003}.  These fields have been estimated
to be a few $10^{10}-10^{12}$\,G, thus consistent with the fields
inferred in classical pulsars. I show in the left panel of
Fig.\,\ref{cyclotron_humps} the X-ray spectrum of X0115+63 and in the
right panel, for comparison, the infrared spectrum of the accreting
magnetic white dwarf in the MCV VV\,Puppis showing cyclotron emission
features \citep{Vis1979}.

The millisecond pulsars (MSPs) are very rapidly spinning neutron stars
with rotational periods $\sol 30$\,ms and magnetic fields typically
$\lse 10^9$\,G. The most widely accepted theory regarding the origin
of MSPs (mostly found in binary systems with white dwarfs or
substellar mass companions) is that they are old neutron stars that
have been spun up (recycled) via mass accretion \citep[as first
suggested by][]{Backus1982}. However, \citet{Ferrario2007} and
\citet{Hurley2010} have demonstrated through population synthesis
calculations that the birthrates of binary MSPs via accretion-induced
collapse (AIC) of white dwarfs can be as large as, and possibly greater
than, those for core collapse. In addition, AIC pulsars can better
reproduce the orbital period distributions of some classes of binary
MSPs.

At the very high end of the field distribution we find the soft gamma
repeaters (SGRs) and the anomalous x-ray pulsars (AXPs), commonly
referred to as the magnetars. They are generally characterised by very
high fields of a few $10^{13}-10^{15}$\,G (but not always, see below)
and spin periods between 2-12\,s which are much longer than those of
radio-pulsars.  The lifetime of magnetars is only a few 10,000 years
and they are often found still embedded in their supernova
remnants. It is still not clear whether these objects were born
rotating very slowly or have spun down rapidly. The persistent and strong
X-ray emission of magnetars ($\sim 10^{35}$\,erg\,s$^{-1}$) is too
large to be powered by their rotational energy. In addition they suffer
from violent bursts lasting $0.1-40$\,s with peak luminosities of up
to $\sim 10^{43}$\,erg\,s$^{-1}$. SGRs also exhibit giant flares with
an energy output of up to $\sim 10^{47}$\,erg\,s$^{-1}$ lasting about
one second. These activities have been attributed to the decay and
instabilities of their magnetic fields \citep{Thompson1996}.  Two very
comprehensive reviews on the properties of magnetars can be found in
\citet{Turolla2015Review} and \citet{Mereghetti2015}.

\begin{figure}
\centerline{\includegraphics[width=0.75\textwidth,clip=]{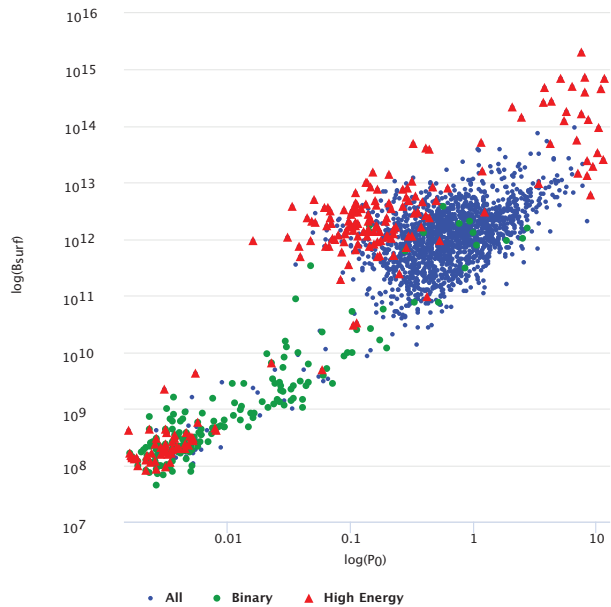}}
\caption{Plot of the surface magnetic field strength $B_{\rm surf}$
  against spin period $P_0$ for pulsars obtained from the ATNF
  catalogue \citep{Manchester2005}.}
\label{pulsars}
\end{figure}

We show in Fig\,\ref{pulsars} the $P-$ diagram of neutron stars
\citep{Manchester2005}.

Another class of neutron stars is that formed by the very quiet and
(probably) weakly magnetic ($\le 10^{10}$\,G) central compact objects
(CCOs), found in young supernova remnants ($<10^4$\,yr).  These are
characterised by: (i) thermal X-ray emission ($0.4$\,keV blackbody
with dimensions of less than 1\,km); (ii) absence of optical or radio
emission; and (iii) lack of a pulsar wind nebula.  Of the known
supernova remnants within 5\,kpc, 14 harbour normal radio pulsars, 6
harbour CCOs, and 1 hosts an AXP giving an incidence of CCOs of about
30\% \citep{Deluca2008}. Thus, the birthrate of CCOs may be similar to
that of normal radio pulsars.  It is still not clear whether these
objects were born with a low field \citep[the proposal of][that CCO
pulsars anti-magnetars]{Gotthelf2008} or whether the field is
submerged due to fallback matter from an accretion disc.
\citet{Vigano2012} have shown that some rather moderate accretion of
$<<0.1$\,M$_\odot$ can bury fields of a few $10^{12}-10^{14}$\,G that
may re-emerge on a time-scale of $10^4-10^5$\,yrs and convert a CCO
into a normal radio pulsar (or even a magnetar). \citet{Popov2012}
suggested that these pulsars would be injected into the general
population at periods of $0.1-0.5$\,s and would be expected to have a
negative braking index because their magnetic fields would be
increasing. They note that this is indeed the case for $20-40$ known
pulsars.

As it is for the magnetic white dwarfs, the origin of magnetic fields
in neutron stars is still unclear. The convection-driven dynamo of
\citet{Thompson1993}, which was inspired by solar dynamo calculations
scaled to neutron stars' physical conditions, predicted fields of up
to $10^{16}$\,G and it is still the most widely accepted theory for
the origin of neutron stars' magnetic fields.  Another avenue for
field generation is given by the differential-rotation driven dynamo
mechanism where toroidal and poloidal fields grow together until they
reach equilibrium values \citep{Braithwaite2006}.  Differential
rotation can also be the outcome of stellar merging events, as
suggested by \citet{Tout2008} and \citet{WTF2014} in the context of
magnetism in white dwarfs. Magnetic fields in neutron stars could also
be explained according to the fossil field hypothesis
\citep{Ferrario2006,Ferrario2008} which invokes magnetic flux
conservation from the main-sequence (or earlier phases) to the compact
star stage.

It is curious that neutron stars with similar dipolar field strengths
(as inferred from their $P$ and $\dot P$) can exhibit a very different
array of emission behaviour. For instance, X-ray observations of
SGR\,0418+5729 by \citet{Rea2010_LowFieldSGR} have revealed a dipolar
magnetic field of only $7.5\times10^{12}$\,G which is typical of
classical radio-pulsars. This indicates that a strong dipolar magnetic
field is not necessary for a neutron star to display the violent
emission characteristics of a magnetar. Instead, these could be caused
by the decay of a large internal toroidal field that does not take
part in the spin-down of the star \citep{Thompson1996, Ferrario2008,
  Rea2010_LowFieldSGR}. It is this toroidal field that could be the
differentiating factor among neutron stars of similar $P$ and
$\dot P$.  Interestingly, SGR\,0418+5729 is located at a rather high
Galactic latitude and its $P$ and $\dot P$ indicate that it is close to
the death line for radio pulsars. Thus, as suggested by
\citet{Rea2010_LowFieldSGR}, this SGR is much older than the other
magnetars, further supporting the hypothesis that its emission and
bursting characteristics \citep[occurring when magnetic stresses
overpower the rigid elastic crust causing crust-quakes][]{Lander2015}
may be due to the reservoir of energy amassed in its super-strong
toroidal field that is slowly dissipating.

The population synthesis calculations of \citet{Ferrario2008}
suggested that massive Of?p stars could be the progenitors of the
magnetars because many of them have been associated with young
clusters of massive stars. On the other hand, it may even be possible
that the stellar merging hypothesis proposed for the explanation of
magnetism on the upper main sequence and in the HFMWDs may also be
applicable to the magnetars. This would give a unified origin for the
fields in most magnetic stars \citep{WTF2014}.

An interesting suggestion is that magnetars could also be responsible
for short and long gamma-ray bursts \citep[GRBs;][]{Turolla2015}. For
instance, \citet{Tout2011} propounded that the merging of an
oxygen-neon white dwarf with the carbon-oxygen core of a naked helium
star during a common envelope phase would produce a rapidly spinning
magnetar giving rise to long GRBs. However, because the birth rate of
magnetars is much higher than that of LGRBs, not all magnetars can be
linked to LGRBs and thus the majority of magnetars is expected to
originate from single-star evolution.

\section{Conclusions}

A 5-10\% incidence of magnetism in stars is observed at all
evolutionary phases, from pre-main-sequence to the compact star stage.
Are these magnetic fields of fossil origin? Taken at face value, the
observational results seem to support this hypothesis.  However, the
total absence of HFMWDs paired with non-degenerate companions in
detached binaries has shed some serious doubts on this theory. The
alternative scenario that would allow us to overcome the problem
presented by this lack of duplicity is that HFMWDs could originate 
from stars that merge during the common envelope phase or from two
merging white dwarfs (double degenerate mergers). Those systems that
survive the common envelope evolution and emerge as close binaries
just before the onset of accretion will evolve into MCVs. A similar
merging scenario could apply to magnetic pre-main-sequence stars, thus
explaining the dearth of short-period binaries among Ap/Bp stars.

The origin of fields in neutron stars is more difficult to ascertain,
partly because of our incomplete knowledge of their magnetic field
strength, structure and evolution. Furthermore, many neutron stars
seem to share the same location in the $P-\dot P$ diagram, thus
suggesting that magnetic field strength alone cannot determine the
observed emission behaviour of neutron stars. One proposal is that a
hidden (internal) ultra-strong toroidal field that is slowly
dissipating could be responsible for most magnetar-like activities
observed in neutron stars.

An interesting possibility that has not been fully explored yet is
that the fields of magnetars could also have originated through binary
interaction, like the mechanism proposed to explain the origin
of fields in the progenitors of the Ap/Bp stars and in the HFMWDs.
The beauty of this scenario is that it leaves a unified picture for
the origin of the highest magnetic fields in all types of stars.

\acknowledgements
L.F. acknowledges support from the Grant Agency of the Czech Republic
(15-15943S) and the organisers of the conference ``Stars with a
stable magnetic field: from pre-main sequence to compact
remnants'' held in Brno. L.F. also wishes to thank all conference
participants for stimulating discussions and in particular Ernst
Paunzen, Martin Netopil, Greg Wade, John Landstreet, Stefano Bagnulo, Steph\'ane
Vennes, Adela Kawka and Alfio Bonanno. 

\bibliography{Ferrario_MagneticStars}

\end{document}